\documentclass[aps,twocolumn,groupedaddress]{revtex4}

\usepackage[oztex]{graphicx}
\begin{document}

\preprint{LA-UR-01-1530}

\title{Quasi-chemical
Theory for the Statistical Thermodynamics of the Hard Sphere Fluid}



\author{Lawrence  R.  Pratt}
\affiliation{Theoretical Division, Los Alamos National Laboratory, 
Los Alamos, NM 87545, USA}

\author{Randall A. LaViolette}
\affiliation{Idaho National Engineering and Environmental Laboratory, 
PO Box 1625, Idaho Falls, ID 83415-2208, USA}

\author{Maria A. Gomez}
\affiliation{Department of Chemistry, Vassar College, Poughkeepsie, NY
12603, USA}

\author{Mary E. Gentile}
\affiliation{Department of Chemistry, Vassar College, Poughkeepsie, NY
12603, USA}


\date{\today}

\begin{abstract}
We develop a quasi-chemical theory for the study of packing
thermodynamics in dense liquids.  The situation of hard-core
interactions is addressed by  considering the binding of solvent
molecules to a precisely defined `cavity' in order to assess the
probability that the `cavity' is entirely evacuated. The primitive
quasi-chemical approximation corresponds to a extension of the Poisson
distribution used as a default model in an information theory approach.
This primitive quasi-chemical theory is in good qualitative agreement
with the observations for the hard sphere fluid of  occupancy
distributions that are central to quasi-chemical theories but begins to
be quantitatively erroneous for the equation of state in the dense
liquid regime of $\rho d^3\!>$0.6. How the quasi-chemical approach can be
iterated to treat correlation effects is addressed.  Consideration of
neglected correlation effects leads to a simple model for the form of
those contributions neglected by the primitive quasi-chemical
approximation. These considerations, supported by simulation
observations,  identify a `break away' phenomena that requires special
thermodynamic consideration for the zero (0) occupancy case as distinct
from the rest of the distribution.  A empirical treatment leads to a one
parameter model occupancy distribution that accurately  fits the hard
sphere equation of state and observed distributions.
\end{abstract}

\maketitle


\section{Introduction}
The quasi-chemical theory\cite{Pratt:MP:98,Pratt:ES:99,Hummer:CPR:2000}
is a fresh attack on the molecular statistical thermodynamic theory of
liquids.  It is intended to be specifically appropriate in describing
liquids of genuinely chemical interest.  But, in view of its generality,
the quasi-chemical theory must be developed and tested for its
description of the paradigmatic hard sphere fluid.  In addition to 
the conceptual point, these developments are expected to be helpful in
subsequent applications of the quasi-chemical theory to real solutions.

The foundational virtues of the hard sphere fluid for the theory of
liquids are widely recognized\cite{WCA} and the interest in this system
continues to evolve\cite{Altenberger:96,Robles:98,Yelash:99,Parisi:00}. Recent
developments of the theory of hydrophobic
effects\cite{Pohorille:JACS:90,Pratt:PNAS:92,Palma,Hummer:PNAS:96,%
Garde:PRL:96,Pratt:ECC,Hummer:PNAS:98,Hummer:JPCB:98,Pohorille:PJC:98,%
Pratt:NATO:99,Gomez:99,Garde:99,Hummer:CPR:2000}, in addition the
related quasi-chemical theory, have emphasized again the significance of
packing issues in a realistic molecular description of complex liquids.
This paper studies the hard sphere fluid and develops default
models with utility in recent information theory
approaches\cite{Hummer:PNAS:96,%
Garde:PRL:96,Pratt:ECC,Hummer:PNAS:98,Hummer:JPCB:98,Pohorille:PJC:98,%
Pratt:NATO:99,Gomez:99,Garde:99,Hummer:CPR:2000,Crooks:PRE:97}.  

A compact derivation requires several preliminary results, including
brief specifications of the potential distribution theorem,  of the
expression of chemical equilibrium, and of the quasi-chemical
formulation. Additionally, the notation here is not elsewhere
standardized because these ideas are unconventional.  The plan of the
paper is thus to collect the necessary preliminary results in
Appendix~\ref{aprelim} so that the conceptual argument needn't be
interrupted. Then we derive the new equation of state format, learn what
we can by comparison of the primitive quasi-chemical
approximation with Monte Carlo simulation results, study correlation
contributions to propose an improved equation of state format, and
finally examine how this improved format works.

Interestingly, though the some of these basic considerations are
regarded as `preliminary,' Eqs.~\ref{hsqca} or \ref{qcarule}, and the
formal identification of the equibrium ratio Eq.~\ref{K},  have been
given before and have wide generality.

\begin{figure}
\begin{center}
\leavevmode
\includegraphics[scale=0.7]{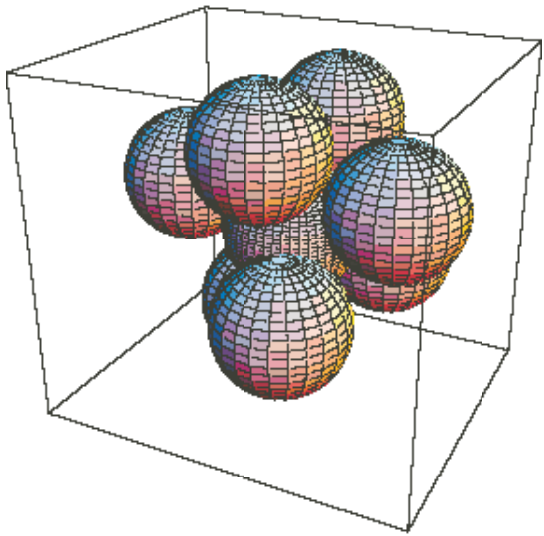}
\end{center}
\caption{An example of an AS$_7$ cluster considered in the text.  The
nucleus (A) is visible in the center.  Each of the ligands (S) overlaps
the nucleus but no other ligand.}
\label{knobs}
\end{figure}

\section{A Quasi-chemical View of the Solvation Free Energy of
Hard Core Solutes}
The preliminary results of Appendix~\ref{aprelim} permit an attack
on the solvation thermodynamics of hard core species built upon a simple
device.   Let's consider a solute A that doesn't interact with the
solvent S molecules at all.  We will consider formation of AS$_n$
complexes and Fig.~\ref{knobs} depicts such a cluster.  The interaction
contribution $\beta\Delta\mu_\mathrm{A}$ is zero and the quasi-chemical
Eq.~\ref{gqca} expresses
\begin{eqnarray}
\ln\left[
\left\langle \left\langle \prod\limits_j {(1-b_{\mathrm{A} j})}
\right\rangle\right\rangle_0  \right] = \ln x_0~.
\label{trick} \end{eqnarray}
But the left side here is a test particle average for solute that
rigidly excludes solvent molecules from the region defined by the
indicator function $b_{\mathrm{A} j}$. If the region is taken as
defining a physically interesting molecule pair excluded volume, then
the right side of Eq.~\ref{trick} gives the negative of the excess
chemical potential for the hard core solute defined by $b_{\mathrm{A}
j}$. This is an example of the well known relation for hard core solutes
$\ln p_0=-\beta\Delta\mu_\mathrm{HC}$ with `HC' denoting `hard
core'\cite{Pohorille:JACS:90,Pratt:PNAS:92,Palma,Hummer:PNAS:96,%
Garde:PRL:96,Pratt:ECC,Hummer:PNAS:98,Hummer:JPCB:98,Pohorille:PJC:98,%
Pratt:NATO:99,Gomez:99,Garde:99,Hummer:CPR:2000}.  This observation
sheds light on the compensation of inner and outer sphere contributions
to the quasi-chemical Eq.~\ref{gqca} but is not surprising. We then
consider `chemical' equilibria for binding of S molecules to the A
molecule. Of course, there is no interaction between the A
molecule and the solvent molecules. The binding is just the occupancy by
a solvent molecules of the `cavity' defined by $b_{\mathrm{A}
j}$. Combining these considerations gives
\begin{eqnarray}
\beta\Delta\mu_\mathrm{HC}  =\ln\left[{1 +  \sum\limits_{m\ge 1} K_m{}
\rho_\mathrm{S}{}^m }\right]~.
\label{hsqca}
\end{eqnarray}
The K$_m$ are well-defined but typically computationally demanding; see
Eq.~\ref{K}. The evaluation of K$_m{}^{(0)}$ will require few-body
integrals over excluded volumes as is discussed in Appendix~\ref{aa}. 
The primitive quasi-chemical approximation is
\begin{eqnarray}
\beta\Delta\mu_\mathrm{HC} \approx \ln\left[1 +  \sum\limits_{m\ge 1} K_m{}^{(0)}
\rho_\mathrm{S}{}^m\lambda{}^m \right]
\label{pqca}
\end{eqnarray}
with $\lambda$ a `mean field' factor that achieves the self-consistency
condition $\sum_n n K_n^{(0)}\rho_\mathrm{S}{}^n\lambda^n =
\rho_\mathrm{S} K_1^{(0)}\sum_n K_n^{(0)}\rho_\mathrm{S}{}^n\lambda^n$. 
This amounts to an extension of the Poisson distribution  for use  in
an information theory procedure\cite{Pratt:NATO:99}.  Here
$\rho_\mathrm{S}$K$_1^{(0)}$=$<$n$>$ is the expected occupancy of the
volume stenciled by $b_{\mathrm{A} j}$.  Thus the multiplicative factors
of $\rho_\mathrm{S}$ in $x_n^{(0)}\propto K_{n}{}^{(0)}\rho
_\mathrm{S}{}^n$ are augmented by a self-consistent `mean field'
$\lambda$\footnote{This point has a twist for the non-profound one
dimensional problem:  This primitive quasi-chemical approximation is
exact for the one dimensional `hard plate' system.  But as a distribution
without evaluation of the mean field factors, this distribution is an
exceedingly weak theory.  Evaluation of the mean field
produces the exact answer because the number of statistical
possibilities is only 2.  The situation for the continuum analog, the
Poisson distribution, is different.  It is not accurate as a
distribution and the same information theory interpretation still gives
an incorrect result for the one dimensional hard plate system.}.

For reuse below, we summarize the technical results of this argument for
hard core solutes, writing
\begin{eqnarray}
\left\langle {\left\langle {e^{-\beta \Delta U_\mathrm{HC} }} \right\rangle }
\right\rangle _0 = {1\over 1 +  \sum\limits_{m\ge 1} K_m{}
\rho_\mathrm{S}{}^m }~. \label{qcarule}
\end{eqnarray}
This combines Eq.~\ref{hsqca} and the potential distribution theorem Eq.~\ref{mu} for
this problem.

\section{Primitive Quasi-chemical Approximation}

\begin{figure}
\begin{center}
\leavevmode
\includegraphics[scale=0.8]{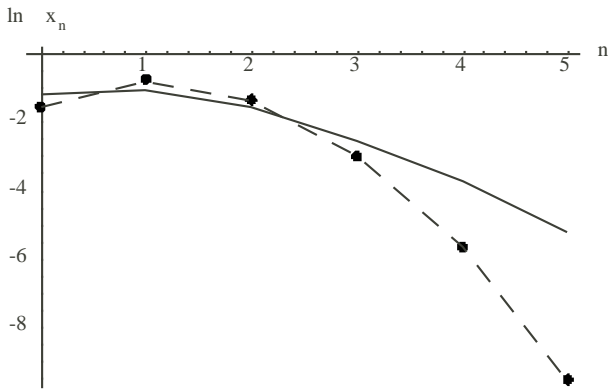}
\end{center}
\caption{For the hard sphere fluid  at $\rho_\mathrm{S}d^3$=0.277,
comparison for n$\le$5 of the Poisson distribution (solid curve) with
primitive quasi-chemical distribution (dashed curve) implemented with
the information theory constraint on the first moment $\sum n
x_n=4\pi\rho_S d^3/3$.  The dots are the results of Monte Carlo
simulation\cite{Gomez:99} as discussed in Appendix~\ref{amc}.  The
primitive quasi-chemical default model depletes the probability of
high-n and low-n constellations and enhances the probability near the
mode.}
\label{compare:fig}
\end{figure}

\begin{figure}
\begin{center}
\leavevmode
\includegraphics[scale=0.8]{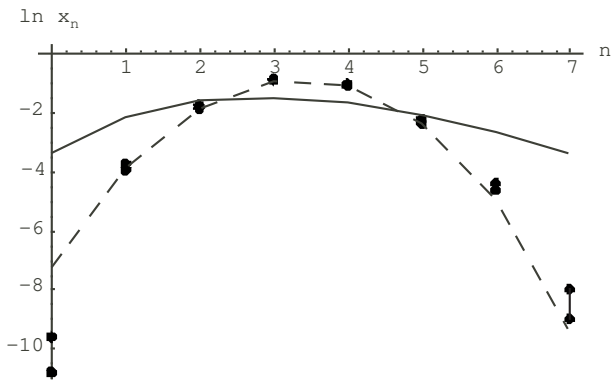}
\end{center}
\caption{As in Fig.~\ref{compare:fig} but for $\rho_\mathrm{S}d^3$=0.8.
The error bars indicate the statistical uncertainty by showing the 67\%
confidence interval.}
\label{compare.04:fig}
\end{figure}

\begin{figure}
\begin{center}
\leavevmode
\includegraphics[scale=0.8]{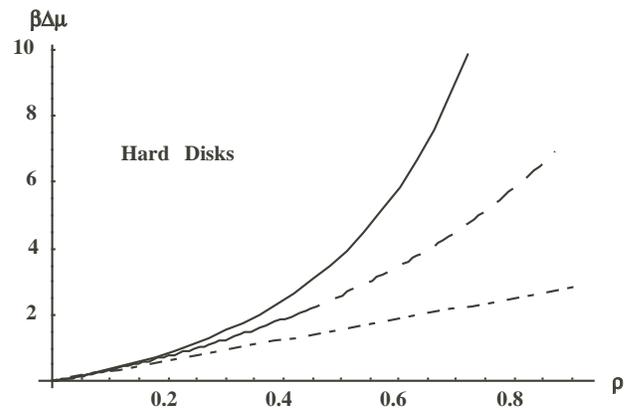}
\end{center}
\caption{$\beta\Delta\mu(\rho)$ for the two dimensional hard disk
fluid on the basis of the primitive quasi-chemical approximation (dashed lined). The 
Ree-Hoover 3,3 Pad\'e approximant\cite{RH:64} is the solid line and
the dash-dot line is the first virial coefficient approximation.}
\label{2dHD:fig}
\end{figure}

\begin{figure}
\begin{center}
\leavevmode
\includegraphics[scale=0.8]{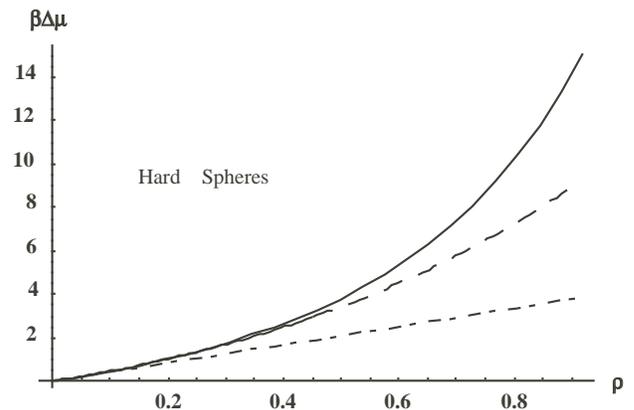}
\end{center}
\caption{$\beta\Delta\mu(\rho)$ for the three dimensional hard sphere
fluid on the basis of the primitive quasi-chemical approximation (dashed
line).  The solid line is the prediction of the Carnahan-Starling
equation of state, taken as the accurate basis for comparison, and
the dash-dot line is the first virial coefficient approximation.}
\label{3dHS:fig}
\end{figure}

We can give a simple demonstration of the quantitative results of the
primitive quasi-chemical theory by considering the hard disk (2d) and
hard sphere fluids (3d). Table~\ref{tab:kn} gives Monte Carlo estimates
of the K$_n{}^{(0)}$ for those cases. The predicted distributions x$_n$
 for two densities are in Figs.~\ref{compare:fig} and
\ref{compare.04:fig}.  Equation of state results
$\beta\Delta\mu(\rho)$ for these systems predicted by this primitive
quasi-chemical theory are shown in Figs.~\ref{2dHD:fig} and
\ref{3dHS:fig}.  The primitive quasi-chemical approximation is
remarkably successfully in all qualitative respects, particularly in
view of its simplicity.  In particular, the predicted occupancy
distributions such as shown in Fig.~ \ref{compare.04:fig} are remarkably
faithful to the data. Nevertheless, the equation of state predictions
begin progressively to incur serious quantitative errors at liquid
densities $\rho d^3\!>$ 0.6, (Fig.~\ref{3dHS:fig}).

\begin{table}
\caption{`Hit-or-Miss' Monte Carlo estimates \cite{HH}, as 
described in Appendix \ref{aa}, of $\ln$K$_n{}^{(0)}$ for unit diameter
hard spheres and disks, respectively.  K$_1{}^{(0)}$=${ 4\pi\over 3}$ ($\pi$) and
K$_2{}^{(0)}$=${17\pi^2\over 36}$ (${3 \sqrt{3}\pi \over 8}$).  The sample size was 24
G-configurations and the results are believed to be accurate to the number
of significant figures given. }
\label{tab:kn}
\begin{tabular}{|c|c|c|}
\hline
n &  spheres (3d) & disks (2d) \\ 
\hline
1 & 1.43241 & 1.14473 \\
2 & 1.53915 & 0.71321 \\
3 & 0.56585 & -1.190 \\
4 & -1.4697 & -5.241 \\
5 & -4.684  & -13.77\\
6 & -9.168  & - \\
7 & -15.46  & - \\
\hline
\end{tabular}
\end{table}

\section{Test of the Equilibrium Ratios} 
\begin{figure}
\begin{center}
\leavevmode
\includegraphics[scale=0.8]{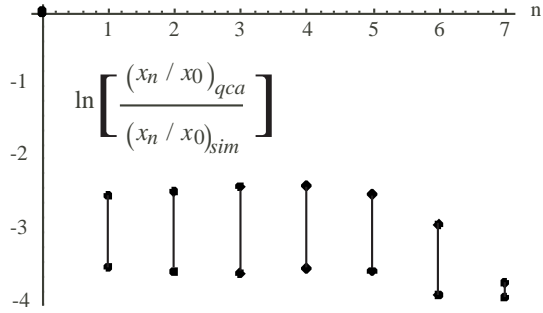}
\end{center}
\caption{$\ln\left[{(x_n/x_0)_{qca} \over (x_n/x_0)_{sim}}\right]$ {\it
vs.} n comparing for the hard sphere fluid the primitive quasi-chemical
approximate populations with those observed by Monte Carlo simulation
for $\rho d^3$=0.8. This normalization focuses on the relative sizes of
K$_n$ and K$_n{}^{(0)}$ suggesting that K$_n\!>$K$_n{}^{(0)}$ even after the
maxent reweighting. The variations are modest {\it except} for the
change between n=0 and n=1.    The error bars indicate the statistical
uncertainty by showing the 67\% confidence interval. In the middle of
the distribution the statistical uncertainty is mostly from the
uncertainty in the denominator factor x$_0$.}
\label{approx.04:fig}
\end{figure}

As a direct check on the primitive quasi-chemical mechanism, we can
focus on testing ideal populations Eq.~\ref{cluster-var.0} as
approximations to formally correct populations Eq.~\ref{cluster-var}. 
It is then natural to consider the ratios $x_j/x_0={K_{n}\rho _\mathrm{S}{}^n}$.  Consideration of
these ratios corresponds to shifting the curves of Figs.~\ref{compare:fig}
and \ref{compare.04:fig} so that the initial point is at the common
value (0,1).   A specific example is shown in Fig.~\ref{approx.04:fig}.
Compared with this normalization, it is clear that the observed
equilibrium ratios K$_n$ are greater than the ideal ratios
K$_n{}^{(0)}$.

\section{Correlations} 

A point of view here is that the geometric weighting   with the
$\lambda$'s of Eq.~\ref{pqca} establishes a mean field that adapts to the
prescribed density.  We now consider how to go beyond that mean field
description. One idea is to extract the features of the summand of
Eq.~\ref{qcarule} that would give purely geometric weighting and then to
analyze what remains.   To this end, we define $\zeta =
\exp(\beta\Delta\mu)$ and consult the formal identification of the
equilibrium ratios Eqs.~\ref{K}.  Thus we can rewrite Eq.~\ref{qcarule}
as
\begin{eqnarray}
\zeta = 1 + \sum_{m\ge 1} x_{0/m}K_m{}^{(0)}\zeta^m \rho_\mathrm{S}{}^m~,
\label{f-exact}
\end{eqnarray}
with 
\begin{eqnarray}
 x_{0/n} \equiv{ \left\langle {\left\langle e^{-\beta \Delta U_n}
\right\rangle } \right\rangle _0 \over \left\langle  e^{-\beta \Delta U}
 \right\rangle _0 }~.
\label{ratio}
\end{eqnarray}
The remarkable Eq.~\ref{f-exact} is formally exact and hasn't been given
before.  The correlation factors $x_{0/m}$ might, in principle, be
investigated on the basis of simulation data and information theory
analysis.  That is likely to a specialized nontrivial activity except of
the lower density cases where the primitive quasi-chemical
approximation is satisfactory.

\subsection{Iterating the Quasi-Chemical Analysis}

Alternatively, the quasi-chemical rules suggest natural
theoretical approximation for the equilibrium ratios given
formally by  Eq.~\ref{K}. Applying
the rule Eq.~\ref{qcarule}, for n$>$0,
\begin{eqnarray}
K_n  & = & \left\langle {\left\langle e^{-\beta \Delta U_n}
\right\rangle } \right\rangle _0 K_n{}{}^{(0)}\zeta^{n+1} \nonumber \\
 & = & { K_n{}^{(0)}\zeta^{n+1} \over 1 + \sum\limits_{m\ge 1} K_{m/n}
 \rho_\mathrm{S}{}^m  } \nonumber \\
 & \approx & { K_n{}^{(0)}\zeta^{n+1} \over 1 + \sum\limits_{m\ge 1}
 K_{m/ n}{}^{(0)} \rho_\mathrm{S}{}^m  }~.
\label{iterate}
\end{eqnarray}
The  K$_{m/n}$ can be understood by considering the chemical equilibrium
\begin{eqnarray} \mathrm{AS'_{n}S_{m=0}\ +\ mS \rightleftharpoons AS'_{n}S_m}~,
\label{rereaction} \end{eqnarray}
{\it i.e.\/} the original $\mathrm{AS'_{n}}$ cluster is the solute and it
provides a nucleus for a constellation of m S particles, different in type
for the S$'$ species.   How to address the calculation of the  K$_{m/ n}{}^{(0)}$
is discussed in Appendix~\ref{ab}.

It is still helpful to focus on the populations even though more coefficients
are involved now.   To do this we consider 
\begin{eqnarray}
\zeta =1+\zeta \sum\limits_{m\ge 1} {\left\langle {\left\langle
{e^{-\beta \Delta U_m}} \right\rangle } \right\rangle _0K_m{}^{(0)}\rho_\mathrm{S}{}
^m\zeta ^m}
\end{eqnarray}
and, to accomodate the additional factor of $\zeta$ multiplying
the terms m$\ge$1, rearrange  so that
\begin{eqnarray}
\zeta =1+{{\sum\limits_{m\ge 1} {\left\langle {\left\langle {e^{-\beta
\Delta U_m}} \right\rangle } \right\rangle _0K_m{}^{(0)}\rho_\mathrm{S}{} ^m\zeta ^m}}
\over {1-\sum\limits_{m\ge 1} {\left\langle {\left\langle {e^{-\beta
\Delta U_m}} \right\rangle } \right\rangle _0K_m{}^{(0)}\rho_\mathrm{S}{} ^m\zeta
^m}}}~.
\label{iqca}
\end{eqnarray}
This last equation is significant particularly because it suggests that a principal
consequence of correlations can be a uniform reweighting of all coefficients
m$\ge$1.  Strikingly, that is
exactly the suggestion of Fig.~\ref{approx.04:fig}.

We can use this insight to push the argument further:  the fact that the
{\em primitive} quasi-chemical populations for m$\ge$1 are correct relative to
each other means that the quantities $\left\langle {\left\langle
{e^{-\beta \Delta U_m}} \right\rangle } \right\rangle _0$ are nearly
exponentially dependent on m.  For, in the first place, when the density
is high, almost all the population is in the center of the distribution,
and the Lagrange multipliers are negligibly affect by the relative
reweighting of the m=0 term.  Then the alteration of the original geometric weighting is
literally irrelevant.  In the second place, when the density is
sufficiently low, these correlation factors are nearly unity anyway.
So we can accurately write
\begin{eqnarray}
\zeta \approx 1+A(\rho_\mathrm{S}{})\sum\limits_{m\ge 1}
K_m{}^{(0)}\rho_\mathrm{S}{}^m\lambda{}^m~. \label{iqca.1}
\end{eqnarray}
x$_0$ `breaks away' from the rest of the distribution and
requires individual consideration when the density is high enough that x$_0$ is sufficiently small due to 
correlation effects.  Nevertheless
\begin{eqnarray}
A(\rho_\mathrm{S}{}) \approx  {\zeta -1 \over \zeta_0 -1}~,
\label{fit-eq}
\end{eqnarray}
where $\zeta_0$ is the primitive quasi-chemical approximate value. Thus,
when the primitive quasi-chemical approximation is sufficiently
accurate, the difficulty of evaluating the corrections should be much
reduced.

\begin{figure}
\begin{center}
\leavevmode
\includegraphics[scale=0.77]{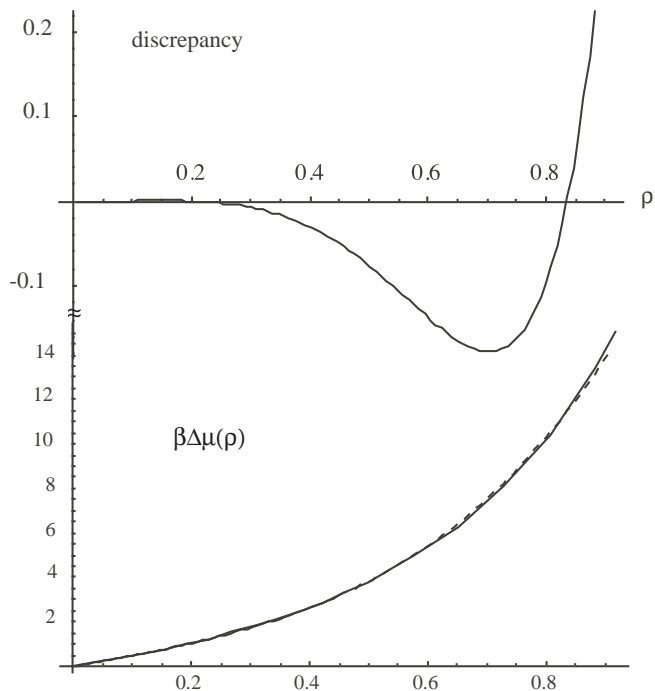}
\end{center}
\caption{Equation of state with the empirical correlation model,
Eq.~\ref{supertheory}.  (Lower panel) The solid line is the
Carnahan-Starling equation of state and the dashed line is the model of
Eq.~\ref{supertheory}. (Upper panel) Discrepancy: the empirical
correlation model (Eq.~\ref{supertheory}) less the Carnahan-Starling
value. The mean absolute discrepancy against Carnahan-Starling equation
of state is about 1\% and the maximum discrepancy is less than 3\%,
nearly as good conformance to the Carnahan-Starling model as that model
to simulation data.  When the final empirical parameter was fitted using
only $\rho_\mathrm{S}d^3\le$ 0.3, the mean absolute discrepancy hardly
changed but the maximum discrepancy doubled. }
\label{dis3d:fig}
\end{figure}

Though it would be interesting to calculate correlation corrections on
the basis of Eq.~\ref{iterate} and Appendix~\ref{ab}, a simpler,
empirical approach suffices for our present purposes.  This is because
the discrepancies seen in Fig.~\ref{3dHS:fig} are substantial but not
problematic and, therefore, the required A($\rho_\mathrm{S}{}$) is
simple.  In particular, the form
\begin{eqnarray}
\beta \Delta \mu =\ln \left( {1+e^{7.361\rho_\mathrm{S}{}
^4}\sum\limits_{m\ge 1} {K_m{}^{(0)}\rho_\mathrm{S}{}^m\lambda{}^m}}
\right)  \nonumber  \\
\label{supertheory}
\end{eqnarray}
conforms accurately to the Carnahan-Starling equation of state;  see
Fig.~\ref{dis3d:fig}.  The literal coefficient in Eq.~\ref{supertheory}
was obtained by fitting on the basis of the Eq.~\ref{fit-eq} to minimize
 the discrepancy with the Carnahan-Starling equation of state.

\begin{figure}
\begin{center}
\leavevmode
\includegraphics[scale=0.8]{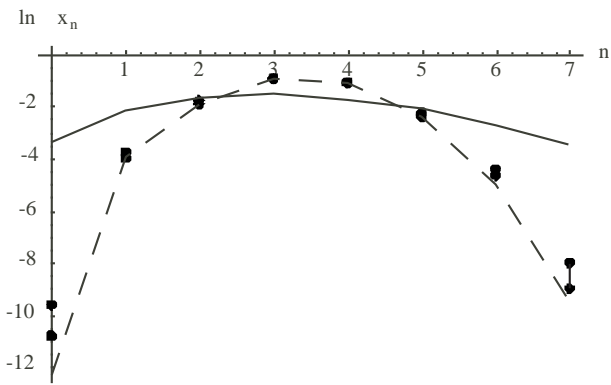}
\end{center}
\caption{Predicted occupancies with the empirical correlation model,
Eq.~\ref{supertheory}, for $\rho_\mathrm{S}d^3$=0.8. Compare to
Fig.~\ref{compare.04:fig}.  Again the solid line is the Poisson
distribution.}
\label{iqca.04:fig}
\end{figure}

The occupancies predicted by this empirical model are depicted in
Fig.~\ref{iqca.04:fig}.

\section{Concluding Discussion}
Our first goal was to work-out how the quasi-chemical
theory, a fresh attack on the the statistical thermodynamic theory of
liquids, applies to the paradigmatic hard sphere fluid.  The second goal
was to work-out theoretical approximation procedures that might assist
in describing dense liquids of non-spherical species.  The new
fundamental results here apply generally to `hard core' molecular models.
The primitive quasi-chemical approximation, the procedure for
iterating the quasi-chemical analysis, and the recognition of the `break away'
phenomenon of Fig.~\ref{approx.04:fig} are likely to be helpful in
understanding packing in dense molecular liquids, beyond the hard
sphere fluid. For the hard sphere system specifically, we have obtained
a simple occupancy model, Eq~\ref{supertheory}, that is likely to be
helpful in a variety of other situations.

One  situation is the description of packing restrictions
when the quasi-chemical theory is used to treat genuinely chemical
interactions, for example in the study of hydration of atomic ions in
water\cite{Rempe:JACS:2000,Rempe:FPE:2001}.  The issue of `context
hydrophobicity' associated with many molecular solutes, including
molecular ions, in water can also be addressed on the basis of
quasi-chemical calculations and the
developments here.

Another situation of current interest is the theory of primitive hydrophobic effects
that has recently been
reborn\cite{Pohorille:JACS:90,Pratt:PNAS:92,Palma,Hummer:PNAS:96,%
Garde:PRL:96,Pratt:ECC,Hummer:PNAS:98,Hummer:JPCB:98,Pohorille:PJC:98,%
Pratt:NATO:99,Gomez:99,Garde:99,Hummer:CPR:2000}. An historical view has
been that the initial issue of hydrophobic effects was the hydration structures and
thermodynamics following from volume exclusion by nonpolar
molecules in liquid water.  The balance of attractive forces that 
might produce drying phenomena was a
secondary concern, except that `drying' was always present in the scaled particle
models\cite{Stillinger:73}.  With the convincing clarification of
the first of these
problems\cite{Pohorille:JACS:90,Pratt:PNAS:92,Palma,Hummer:PNAS:96,%
Garde:PRL:96,Pratt:ECC,Hummer:PNAS:98,Hummer:JPCB:98,Pohorille:PJC:98,%
Pratt:NATO:99,Gomez:99,Garde:99,Hummer:CPR:2000}, the issue of
drying phenomena has been now taken up more
enthusiastically\cite{Hummer:98,LCW,Hummer:CPR:2000}.  In this context, we note
that the striking success of the two-moment information models and the
Pratt-Chandler
theory\cite{Pohorille:JACS:90,Pratt:PNAS:92,Palma,Hummer:PNAS:96,%
Garde:PRL:96,Pratt:ECC,Hummer:PNAS:98,Hummer:JPCB:98,Pohorille:PJC:98,%
Pratt:NATO:99,Gomez:99,Garde:99,Hummer:CPR:2000,PC} is due, in part, to
a fortuitous balance of a `gaussian' approximation in the
theory\cite{Pratt:NATO:99} and a compensating disregard for
drying possibilities\cite{Pratt:80}; both of these compensating approximations are
expected to be benign for small molecule solutes\footnote{Another factor is that the dimensionless densities
relevant for the first hydrophobic applications aren't as large as the
most difficult cases considered here.}.  One ingredient in a
better understanding of this situation is a careful solution to the case
where drying phenomena are entirely absent.  That ingredient is better
in hand with the results above.

\begin{acknowledgments}This work was supported by the US Department of
Energy under contract W-7405-ENG-36 and the LDRD program at Los Alamos.
LA-UR-01-1530. Work at the INEEL was supported by the Office of
Environmental Management, U.S. Department of Energy, under DOE-ID
Operations Office Contract DE-AC07-99ID13727.
\end{acknowledgments}

\appendix

\section{Preliminary Results}\label{aprelim}
\subsection{Potential Distribution Theorem}
The potential distribution theorem\cite{Pratt:ECC,Pratt:MP:98,Pratt:ES:99,Widom:JPC:82}
may be expressed as 

\begin{eqnarray}
\rho _\sigma = \left\langle {\left\langle {e^{-\beta \Delta U}}
\right\rangle } \right\rangle _0z_\sigma \left( {q_\sigma /V} \right)
\label{double}
\end{eqnarray}
where $\rho_\sigma$ is the density of molecules of type
$\sigma$ (the `solute' under consideration), $z_\sigma =
\exp(\beta\mu_\sigma)$ is the absolute activity of that species,
q$_\sigma$ is the single molecule partition function for that species,
and V is the volume.  The double brackets
$\left<\left<\ldots\right>\right>_0$ indicate the average over the
thermal motion of the solute {\em and} the solvent under the conditions
of no interaction between them, and the averaged quantity is the
Boltzmann factor of those interactions. The average indicated here is
the ratio of the activity of an isolated solute, $\rho_\sigma
V/{q_\sigma}$, divided by the absolute activity, $z_\sigma$, of the
actual solute.  Thus
\begin{eqnarray}
\beta\mu_\sigma = \ln \left[{V \rho _\sigma \over \left\langle
{\left\langle {e^{-\beta \Delta U}} \right\rangle } \right\rangle _0
q_\sigma }\right]~.
\label{mu}
\end{eqnarray}
This is a formal result to the extent that evaluation of the quantities
on the right side typically will  involve nontrivial calculations on
many-body systems.

\subsection{Chemical Equilibrium}
The traditional chemical thermodynamic consideration of a chemical transformation such as
\begin{eqnarray}
\mathrm{n_\mathrm{A} A + n_\mathrm{B} B \rightleftharpoons  n_\mathrm{C} C + n_\mathrm{D} D }
\end{eqnarray}
with the formal result of Eq.~\ref{mu} leads to the formal expression
\begin{eqnarray}
K & \equiv & {\rho _\mathrm{C}{}^{n_\mathrm{C}} \rho _\mathrm{D}{}^{n_\mathrm{D}}\over \rho _\mathrm{A}{}^{n_\mathrm{A}} \rho _\mathrm{B}{}^{n_\mathrm{B}} } \nonumber \\
& = &
{
\left(\left\langle\left\langle e^{-\beta \Delta U_\mathrm{C}} \right\rangle \right\rangle _0
{q_\mathrm{C}\over V} \right){}^{n_\mathrm{C}} 
\left(\left\langle\left\langle e^{-\beta \Delta U_\mathrm{D}} \right\rangle \right\rangle _0
{q_\mathrm{D}\over V} \right){}^{n_\mathrm{D}}\over 
\left(\left\langle\left\langle e^{-\beta \Delta U_\mathrm{A}} \right\rangle \right\rangle _0
{q_\mathrm{A}\over V} \right){}^{n_\mathrm{A}} 
\left(\left\langle\left\langle e^{-\beta \Delta U_\mathrm{B}} \right\rangle \right\rangle _0
{q_\mathrm{B}\over V} \right){}^{n_\mathrm{B}} 
}~.
\label{K}
\end{eqnarray}
This should be compared to the textbook result for ideal gas
systems\cite{mcq}. That comparison shows that the single molecule
partition functions are multiplicatively augmented by the test particle
averages\footnote{A more conventional view of these results is that the
concentrations are multiplicatively augmented by the test particle
averages and that serves to identify activity coefficients.  That point
of view has no special utility here.}.  But otherwise the structure of
this important result is unchanged. The conclusion here is that the
equilibrium ratios are well-defined objects though formal to the extent
that nontrivial computational effort would be required to evaluate them
on the basis of molecular information.

\subsection{Quasi-chemical Theory}
The quasi-chemical develop starts from consideration of a distinguished
molecule in the solution and seeks to evaluate the chemical potential on
the basis of events occurring within a defined `inner sphere.' For a
species of type A, that definition is codified by specifying a function
$b_{\mathrm{A} j}$ that is equal to one (1) when solution molecule j is
inside the defined region and zero (0) otherwise. Our starting point
can be \cite{Pratt:ES:99}
\begin{eqnarray}
\beta\Delta\mu_\mathrm{A} &  = & \ln x_0  \nonumber \\
& - & \ln\left[
\left\langle\left\langle {e^{-\beta
\Delta U_\mathrm{A}}}\prod\limits_j {(1-b_{\mathrm{A} j})}
\right\rangle\right\rangle_0  \right],
\label{gqca} \end{eqnarray}
where $x_0$ is the fraction of A solute species with zero (0) neighbors
in the defined region.  $\Delta U_\mathrm{A}$ is the interaction energy
of the solvent with the solute A that is treated as a test particle.
The potential distribution theory perspective on Eq.~(\ref{gqca}) is
\begin{eqnarray}
x_0 & = & \left\langle \prod\limits_j {(1-b_{\mathrm{A} j})} \right\rangle
\nonumber \\ 
& = &{ \langle\langle{e^{-\beta \Delta U_\mathrm{A}}}\prod\limits_j
{(1-b_{\mathrm{A} j})}\rangle\rangle_0 \over \langle\langle {e^{-\beta \Delta U_\mathrm{A}}}
\rangle\rangle_0} .
\label{x0}
\end{eqnarray}

The first, or {\em chemical} term, of Eq.~\ref{gqca} can be analyzed with chemical
concepts associated with the reactions
\begin{eqnarray} \mathrm{AS_{n=0}\ +\ nS \rightleftharpoons AS_n}
\label{reaction} \end{eqnarray}
Here the indicated complexes are
composed of $n$ solvent (S) molecules within the defined
region.  Remember that the
A molecule is a `distinguished' solute molecule considered at the
lowest non-zero concentration \cite{Pratt:MP:98}. The fractional amount of
A species with a given solvation number $n$ can be described by a
chemical equilibrium ratio
\begin{eqnarray}
K_n={\rho_\mathrm{AS_n} \over \rho_\mathrm{AS_{n=0}}
\rho_\mathrm{S}{}^n }~.
\label{Kn-ob}
\end{eqnarray}
The $\rho_\sigma$ are the number densities and, in particular,
$\rho_\mathrm{S}$ is the bulk number density of solvent molecules since the
distinguished A molecule is infinitely dilute. This notation
permits the normalized re-expression
\begin{eqnarray}
x_n={{K_{n}\rho _\mathrm{S}{}^n} \over {1 + \sum\limits_{m\ge 1}
{K_{m}\rho _\mathrm{S}{}^m}}} .
\label{cluster-var}
\end{eqnarray}
Since this yields
\begin{eqnarray}
x_0{}^{-1}={1 + \sum\limits_{m\ge 1}
{K_{m}\rho _\mathrm{S}{}^m}} ,
\label{x0.f}
\end{eqnarray}
the original Eq.~(\ref{gqca}) can be re-expressed as 
\begin{eqnarray}
\beta\Delta\mu_\mathrm{A}  &=&  -\ln \left[1 +  \sum\limits_{m\ge 1}
{K_{m}\rho _\mathrm{S}{}^m}\right] \nonumber\\&&- \ln\left[
\left\langle\left\langle {e^{-\beta
\Delta U}}\prod\limits_j {(1-b_{\mathrm{A} j})}
\right\rangle\right\rangle_0  \right].
\label{gqca.f} \end{eqnarray}
The virtue of these rearrangements is that the natural first approximation is
\begin{eqnarray}
x_n\approx x_n{}^{(0)} = {{K_{n}{}^{(0)}\rho _\mathrm{S}{}^n} \over
{1 + \sum\limits_{m\ge 1} {K_{m}{}^{(0)}\rho _\mathrm{S}{}^m}}} .
\label{cluster-var.0}
\end{eqnarray}
 The K$_{n}{}^{(0)}$ are equilibrium ratios for the chemical reaction
(\ref{reaction}) in an ideal gas.  This formulation and the
approximation of Eq.~(\ref{cluster-var.0}) are closely related
\cite{Pratt:ES:99} to the quasi-chemical (or cluster-variation)
approximations of Guggenheim \cite{Guggenheim:35}, Bethe
\cite{Bethe:35}, and Kikuchi \cite{brush}.

This approach should have greatest utility where the chemical balances of
Eq.~\ref{reaction} are dominated by inner sphere chemistry that can be
captured with computations on clusters.  Such chemical interactions are
often much larger than the {\em outer sphere\/} contribution, the
right-most term of Eq.~\ref{gqca}.  

But that outer sphere contribution remains and can't be neglected
forever.  An interesting example based on simulation of liquid water was
discussed recently\cite{Hummer:CPR:2000}.  There the x$_0$ was estimated
from molecular dynamics results and the remainder, the outer-sphere
contributions to $\beta\Delta\mu$, were positive, suggesting domination
of those outer-sphere contributions  by the packing constraints studied
here. A principal goal of the present work is the development of a
reasonable approach for describing the packing issues necessary for
treating those outer sphere contributions.

Reiss and Merry\cite{Reiss:81} analyzed population relations analogous to
Eq.~\ref{cluster-var} but with activities appearing in the place of
densities and with coefficients, here the equilibrium ratios K$_n$,
appropriately different.  The additional formal point here is the
replacement of the activity by the density that permits the
identification of the K$_n$ in Eq.~\ref{hsqca}, and then further permits
consideration of the mean field treatment Eq.~\ref{pqca} on the basis of an
information theory constraint when  K$_n{}^{(0)}$ will be used.  At this
stage, the quasi-chemical approximation achieves a particularly
primitive character and deviates from the goal of bounding these
thermodnamic quantities that was pursued by Reiss and
Merry\cite{Reiss:81}.

\section{Calculation of the K\lowercase{$_n{}^{(0)}$} for hard spheres and hard
disks}\label{aa} The K$_n{}^{(0)}$ sought for reaction
Eq.~\ref{reaction} are given by
\begin{eqnarray}
K_n{}^{(0)}  = 
{ {q_\mathrm{AS_n}}  \over 
\left({q_\mathrm{S}/ V}\right)^n 
{q_\mathrm{AS_{n=0}}}} ~.
\label{K0n}
\end{eqnarray}
(See Eq.~\ref{K}.) For this problem, $ q_\mathrm{S} = V/\Lambda_S{}^3$
with $ \Lambda_S$ a thermal deBroglie wavelength for S but these momentum
integrals cancel perfectly and are irrelevant as usual.  Therefore,
\begin{eqnarray}
n!K_n^{(0)}=\int\limits_A {d^3r_1\ldots \int\limits_A {d^3r_n\left(
{\prod\limits_{j>i=1}^n {e(i,j)}} \right)}}~.
\label{integral}
\end{eqnarray}
The notation $\int_A {d^3r_k}$ indicates the three-dimensional spatial
integral over the volume of the A-ball, a sphere of radius 1.  The
indicated integrand is thus 3n dimensional. The integrand is zero (0) if
$\vert\bf{r}_i-\bf{r}_j\vert<1$ (overlap) for any (ij) and one (1)
otherwise.  Thus the integral can be estimated by sampling n-point
uniform placments  in the A-ball and scoring the fraction of such
placements that are free from overlaps between the n unit diameter
S-spheres.  This approach fails for n larger than those presented in
Table~\ref{tab:kn}. But larger clusters were not observed in our
simulation of the fluid, so our approach should be regarded as
satisfactory.

The analogous two dimensional procedure was used for the hard disk results.

\section{Calculation of the K\lowercase{$_{m/n}{}^{(0)}$}}\label{ab}
In contrast to Appendix \ref{aa}, here the ratio sought is
\begin{eqnarray}
K_{m/n}{}^{(0)}  = 
{ {q_\mathrm{AS_n'S_m}}  \over 
\left({q_\mathrm{S}/ V}\right)^m 
{q_\mathrm{AS_{n}'}}} ~,
\label{K0mn}
\end{eqnarray}
corresponding to the reaction Eq.~\ref{rereaction}. Again, the explicit
factors of V, the momentum integrals, and the factor of $n!$ all cancel
perfectly so that
\begin{eqnarray}
m!K_{m/n}^{(0)}=\left\langle {\int\limits_{AS'_n} {d^3r_1\ldots
\int\limits_{AS'_n} {d^3r_m\left( {\prod\limits_{j>i=1}^m {e(i,j)}}
\right)}}} \right\rangle~.\nonumber \\
\label{ave-int} 
\end{eqnarray}
Here the notation $\int_{AS_n'} {d^3r_k}$ indicates an integral over the
excluded volume of an AS$_n'$ complex to an S ligand.  The SS excluded
volume, the integrand, is the same as before.  But the structure of
the AS$_n'$ complex fluctuates and the volumes obtained for specific
structures are averaged over these fluctuations.  The brackets $\langle
\ldots \rangle$ indicate the average over the structures of the isolated
AS$_n'$ complex. This averaging is permitted and governed by the
non-trivial denominator that appears in Eq.~\ref{K0mn}.

Operationally, the calculation can be much as in Appendix \ref{aa}
except for (a)  averaging utlizing a Metropolis Monte Carlo calculation for
the n ligand spheres in the star AS$_n'$; and (b) random placements of
the m additional points are into a sphere of radius two (2) since that
would fully enclose any conformation of the cluster.

\section{Calculation of \lowercase{x$_{n}$} for the Hard Sphere Fluid}
\label{amc}
 
The probability that there are $\mathrm{n-1}$ points in a sphere of
radius $r$, x$_{n-1}(r)$, can be obtained from the
distribution, $4 \pi \rho_S r ^2\mathcal{D}_{n}(r)$,  of the distance $r$ to the
nth nearest neighbor of an arbitrary point. The probability that there
are no more than $\mathrm{n-1}$ molecules in the void is equal to the
probability that the nth nearest neighbor is at least $r$ away
from the void center
 
\begin{equation}
\sum_{m=0}^{n-1} x_{m}(r) = 4 \pi \rho_S\int_{r}^{\infty} \mathcal{D}_{n}(y)
y ^2 dy ~.
\end{equation}
Isobaric-isothermal Monte Carlo can be used to calculate
$\mathcal{D}_{n}(r)$.  x$_n(r)$ for a
range of $r$ can be obtained from the distributions
$\mathcal{D}_{n}(r)$. To increase the accuracy of the estimated
$\mathcal{D}_{n}(r)$ for rarely observed $r$, small and large, a
specific point in the simulation volume was chosen, and the sampling
probability was reweighted by $\left[4\pi \rho_\mathrm{S}r_1{}^2
e^{-4\pi \rho_\mathrm{S}r_1{}^3/3}+ C\right]^{-1}$ where $r_j$ is the
distance from the chosen point to the jth nearest center for each
configuration and C is an empirically chosen, dimensional constant. This
importance sampling

\begin{equation}
4 \pi \rho_S r^2\mathcal{D}_{n}(r)=
{
\left<\left[4\pi \rho_\mathrm{S}r_1{}^2 e^{-4\pi \rho_\mathrm{S}r_1{}^3/3}+
C\right] \delta(r-r_n)\right>
\over
\left<\left[4\pi \rho_\mathrm{S}r_1{}^2 e^{-4\pi \rho_\mathrm{S}r_1{}^3/3}+
C\right]\right>
}
\label{importance}
\end{equation}
is based upon the idea that $\mathcal{D}_{1}^{(0)}(r)=e^{-4\pi
\rho_\mathrm{S}r^3/3}$ is the function\cite{Chandrasekhar}
appropriate for a random distribution of spheres. This idea attempts to
make the observed distribution of the distance to the nearest particle
more nearly uniform. The constant C was included to avoid an unbounded
weighting function.  The denominator of Eq. D2 is just a normalizing
factor on the distribution. The denominator of Eq.~\ref{importance}
merely provides a normalizing factor.

Isobaric-isotermal ensembles of 108 and 256 hard spheres were
sufficient.  The Carnahan-Starling equation was used to find the $\beta$p
needed for a hard sphere simulation at each specific density.


%
%

%



\end{document}